\documentclass[12pt,a4paper]{article}
\usepackage[utf8x]{inputenc}
\usepackage[margin=2cm]{geometry}
\usepackage{amsmath}
\usepackage{amsfonts}
\usepackage{amssymb}
\usepackage{graphicx}
\usepackage{hyperref}
\usepackage{siunitx}
\usepackage{cite}

\DeclareGraphicsExtensions{pdf,png,jpg}
\graphicspath{{FIGS/}}

\title{Thermodynamics, entropy and waterwheels}
\author{Franco Bagnoli\\
Dept. Physics and Astronomy and CSDC, University of Florence\\
via G. Sansone 1, 50019 Sesto Fiorentino, Italy\\
Also INFN, sez. Firenze\\
\texttt{franco.bagnoli@unifi.it}
}

\def\dbar{{\mathchar'26\mkern-12mu d}}
\catcode`"=\active
\def"#1"{``#1''}

\begin{document}

\maketitle
\abstract{
	In this paper the analogy between a thermal engine and a waterwheel is developed in details, showing that the analogous of the flow of water in an hydraulic engine is the flow of entropy in a thermal one. This analogy mat serve to analyse in details the aspects of a thermal engine that are quite distant from pupils' intuition. The discussion is complemented by an illustration of the working of thermal and hydraulic engines at the microscopic level, thus furnishing a view of what heat and work are at this level, and introducing the statistical and information aspects of entropy. Both hydraulic and thermal non-ideal engines are also discussed. 
}

\section{Introduction}
Students have difficulties in appreciating the concept of entropy~\cite{kesidou}. Indeed, there are  two popular "versions" of this quantity: Clausius' thermodynamic entropy,  $dS=\dbar Q/T$, usually defined by the help of  Carnot engines;  and Shannon's information entropy $S=-\sum_i p_i\log(p_i)$, where I assumed  Boltzmann constant equal to one, i.e.,  temperature measured in energy units.
 
The Boltzmann space-volume entropy $S=\log(\Omega)$ can be considered a special case of Shannon's in the case of a isolated system for which every portion of the accessible phase space is equally probable.  I shall come back on these quantities later, defining them in a proper way. 
 
The thermodynamic definition is indeed quite hard to be taught.  The thermodynamic formalism  does not furnish a clear illustration of the difference between the ordered motion, i.e., the kinetic energy, and the disordered one, or temperature. A common choice is that of avoiding the thermodynamic formulation, insisting on the statistic character of entropy.

For instance, in Ref.~\cite{atkins} the author emphasizes that the work is given by the collective, ordered motion of the microscopic components (for instance, the atoms of
a tennis ball that travels all together with the same velocity) while thermal motion is disordered (the same atoms that chaotically oscillate around their 
equilibrium position). In order to illustrate the concept of disorder, the author uses a cellular automaton, i.e., a  lattice gas model of the universe, 
constituted by a grid of Boolean pixels. The energy corresponds to the "magnetization", i.e.,  the number of "up" spins. By using a kind of long-range spin-exchange stochastic dynamics, the author illustrates the concept of thermal equilibrium between two systems. The temperature concept is introduced by means of the probability distribution of energy.
After that, the author illustrates, using a different model, the dispersal of ordered kinetic energy into heat, and, by means of the original Boolean model, how to build a Carnot engine that converts part of the heat into work. Kincanon~\cite{kincanon} uses a similar approach. 

Styer~\cite{styer} illustrates the qualitative character of entropy as related to the class (distribution) of patterns from which a sample, presented again as a lattice gas, is extracted. Schoepf~\cite{schoepf} uses a system formed by  equally spaced energy levels of almost non-interacting oscillators,  in order to introduce 
the concept of equilibrium distribution and the notion of entropy.

Leff~\cite{Leff1,Leff2,Leff3,Leff4,Leff5} again uses microscopic models in a query-and-answer mode  trying to relate the entropy to the concept of energy spreading, which in my opinion is more related to entropy production, as done in the Kolmogorov-Sinai formulation~\cite{kolmogorov,sinai,ott}. Leff's formulation is anyhow quite interesting and is a useful tool for teachers. 

As we shall also see in Section~\ref{sec:discrete}, the statistical or information formulation of entropy essentially requires that the phase space can be treated as a numerable set of boxes, either intrinsically (quantum approach) or artificially, using a coarse-graining that can be taken order the Planck constant $h$. 

The "microscopic" formulation is indeed capturing, but the beauty of the thermodynamic formulation is that of being independent of any model. I thus think that students should be presented with both approaches. 

The classic derivation of entropy makes use of thermal engines, which are generally outside the direct experience of pupils, who probably see them as "mysterious" objects.

Actually, I think that the biggest source of confusion is the concept of heat, which "enters in" and "exits from" thermal engine, much like a fluid. It is not a coincidence that  Carnot was able to 
derive the second law of thermodynamics working within the caloric scheme~\cite{carnot}.  In his book, he often repeats that the ``fall'' of the caloric through a heat engine is equivalent to the fall of the water through a water wheel.  
\begin{quote}
	The motive power of a waterfall depends on its
	height and on the quantity of the liquid; the
	motive power of heat depends also on the quantity
	of caloric used, and on what may be termed, on
	what in fact we will call, the height of its fall, 
	that is to say, the difference of temperature of the
	bodies between which the exchange of caloric is
	made.  \cite{carnot} p. 61,
\end{quote}
and he uses this analogy for establishing the idea of maximum efficiency
\begin{quote}
	According to established principles at the present
	time, we can compare with sufficient accuracy the
	motive power of heat to that of a waterfall. Each
	has a maximum that we cannot exceed, \dots \cite{carnot} p. 60.
\end{quote}
Indeed, I am pretty sure that, as Carnot, many of our student visualize the abstract concept of a thermal engine as a kind of "heat-wheel". 

Actually, one can play the analogy between thermal and hydraulic machines all the way down, and discover what really "falls" through a thermal engine, i.e., the entropy. The first steps, the analogy between the mass of water and the entropy, is not original, it can be found for instance in Ref.~\cite{baracca}, but I think that it can be further developed.
I shall present both the macroscopic analogy between waterwheel and thermal engines, and the microscopic one. 

This article is not meant to be directly used by teachers, it just contains what I hope are interesting suggestions that have to be further developed. 

The outline of this paper is the following: in Sec.~\ref{sec:thermo} I quickly review the derivation of the entropy in the frame of classic thermodynamics. In the next Section the analogy between thermal engines and hydraulic waterwheels is presented. The microscopic approach to both problems is presented in Sec.~\ref{sec:micro}, and the discussion about non-ideal engines (both thermal and hydraulic) is reported in Sec.~\ref{sec:nonideal}. Finally, possibly abusing of the previous analogy, the third law of thermodynamic for waterwheels is presented in Sec.~\ref{sec:thirdlaw}. Conclusions are drawn in the last Section.

\section{Thermodynamic entropy}\label{sec:thermo}
Let me quickly recall the thermodynamic introduction to the entropy. 

One starts introducing the empirical temperature, by means of some thermometer. The zeroth principle says that the temperature is an intensive quantity, and that this quantity  is the same for any two systems  in thermal contact, i.e., that can exchange energy for a sufficiently long period, and are isolated from the outside world.  

By using gas thermometers, one can notice that in the limit of high rarefaction, all  gases behave in the same way, i.e., that they became "ideal".  The absolute zero can be defined as the extrapolated temperature at which the volume and pressure of an ideal gas  will be zero. Clearly, no gas remains ideal at that temperature. 

\begin{figure}[t]
	\begin{center}
		\includegraphics[width=0.4\columnwidth]{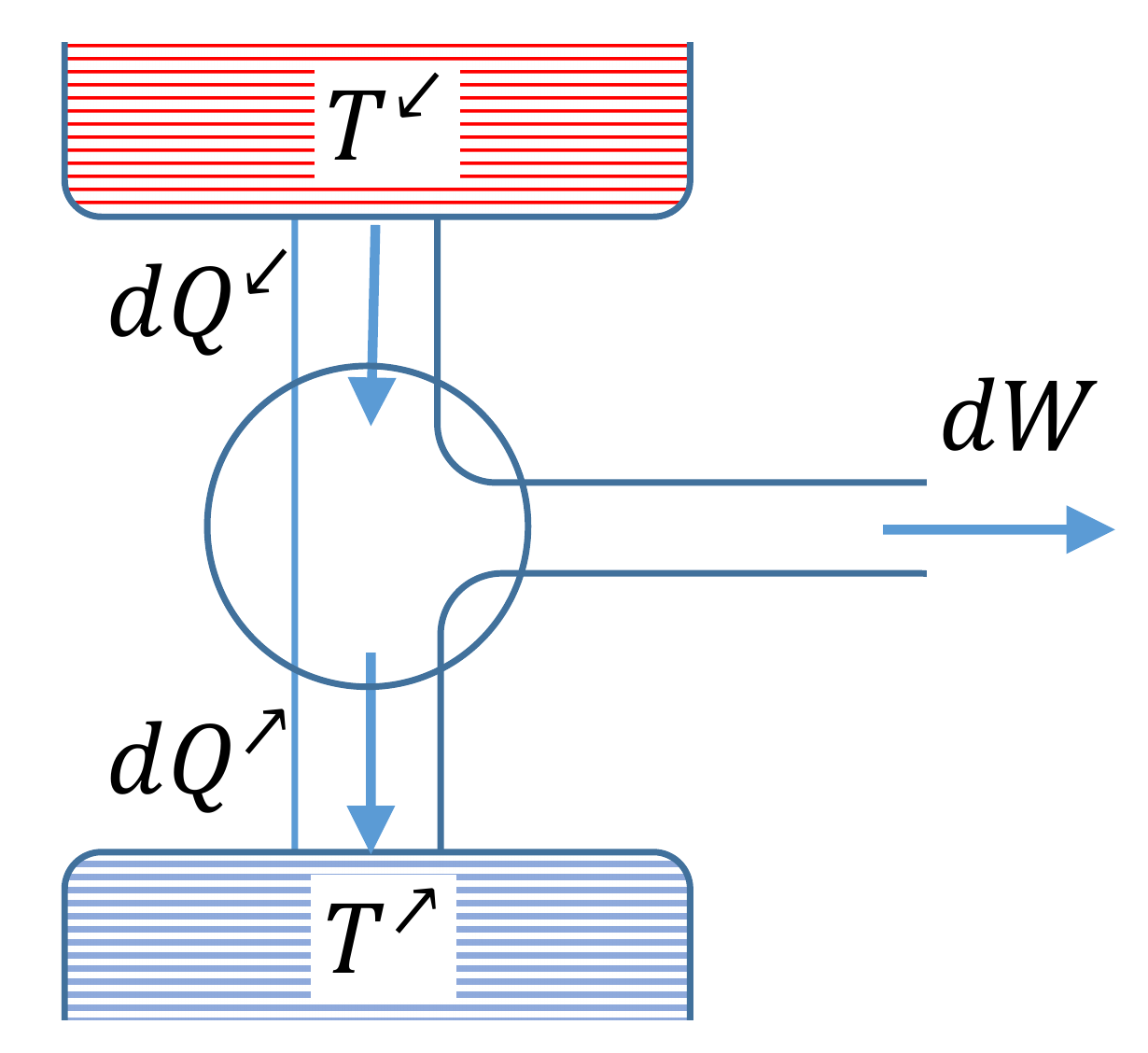}
	\end{center}
	\caption{A schematic thermal engine. The circle represents the cyclic behaviour of the engine.}
	\label{fig:tengine}
\end{figure}

\begin{figure}[t]
	\begin{center}
		\includegraphics[width=0.6\columnwidth]{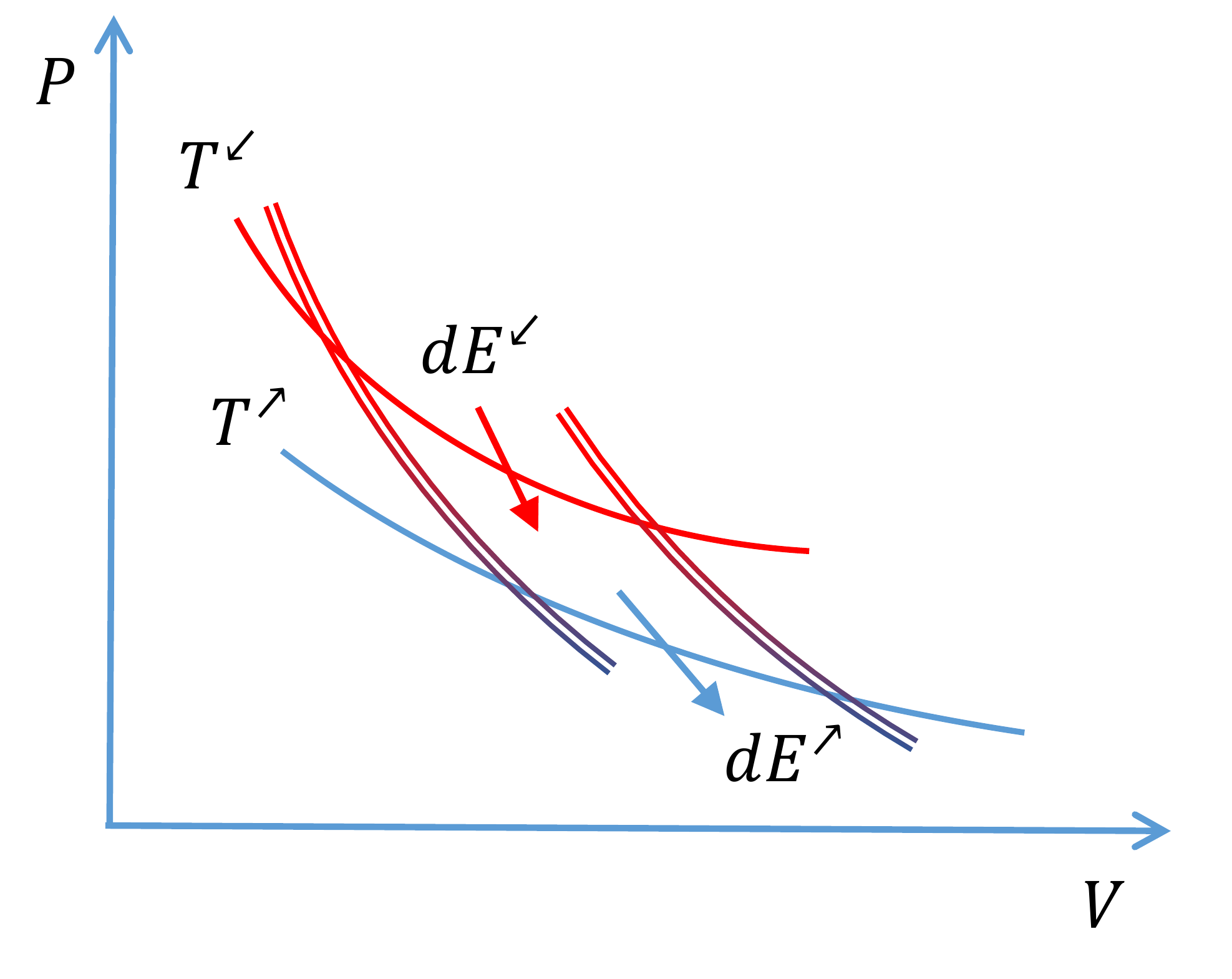}
	\end{center}
	\caption{Carnot cycle.}
	\label{fig:carnot}
\end{figure}

One can now introduce the concept of work $W$ as the mechanical energy exchange, and the heat $Q$ as the energy exchanged by other means, and introduce the conservation of energy (first principle) 
\[
d Q=dU + d W,
\]
which for a cycle ($\Delta U=0$) becomes (see Fig.~\ref{fig:tengine})
\[
W= Q^{\swarrow}-Q^{\nearrow}.
\]
It is time to introduce thermal engines and the Carnot one, defining the efficiency $\eta$:
\[
\eta = \frac{W}{Q^{\swarrow}}=1-\frac{Q^{\nearrow}}{Q^{\swarrow}}.
\]
It is now possible to formulate the Kelvin and Clausius statements of the second law,   showing  that reversible engines have the maximum efficiency, and that any engine operating between two given heat reservoirs has that same efficiency.

By means of the Carnot cycle (Fig.~\ref{fig:carnot}) and the laws of perfect gases, one can easily show that for all reversible engines operating between two heat reservoirs at temperatures $T^{\swarrow}$ and $T^{\nearrow}$, respectively,  the efficiency is
\[
\eta =1-\frac{T^{\nearrow}}{T^{\swarrow}},
\]
i.e., that the Carnot engine can serve as a thermometer, defining the thermodynamic temperature as proportional to the heat exchanged. The Carnot engine also serves for introducing the entropy 
\[
dS=\frac{d Q}{T}
\]
and using the Clausius theorem one arrives at the entropy-based definition of the second law, i.e., that for a closed system the entropy is always increasing. 

Finally, let us consider two systems, $A$ and $B,$ in thermal contact but isolated from the outside world. The total energy $E=E_A+E_B $ is constant,  and the total entropy $S=S_A+S_B$ is given by the sum of the entropy of the two subsystems. If a fluctuation $dE_A$ increases a bit the  energy $E_A$ of the $A$ system, one must have $dE_B=-dE_A$. Now, if the two systems are at equilibrium, the second principle says that 
\[
\frac{dS}{dE_A} = 0.
\]
This implies that 
\[
\frac{dS}{dE_A} =\frac{dS_A}{dE_A} +\frac{dS_B}{dE_A} =  \frac{dS_A}{dE_A} -\frac{dS_B}{dE_B}=0,
\]
and therefore, at equilibrium
\[
\frac{dS_A}{dE_A} =\frac{dS_B}{dE_B}.
\]
Since the zeroth principle says that two systems at equilibrium and in thermal contact have the same temperature $T$, this has to be related to $dS/dE$. In order to make it coincide with that coming from of the ideal gas, one defines
\[
\frac{1}{T} = \frac{dS}{dE}.
\]
In the previous derivation I avoided using partial derivatives, since the notation is clear enough. 

In my opinion, there are three big problems with this presentation. First, Fig.~\ref{fig:tengine} is quite misleading: the heat $dQ$ really looks as a falling fluid (and this may be the reason why in many books this picture is represented in an horizontal way). Secondly, the entropy thus defined has no "physical" attribute. Finally,  it is not absolutely clear what work and heat are at a microscopic level. 

\begin{figure}[t]
	\begin{center}
		\includegraphics[width=0.8\columnwidth]{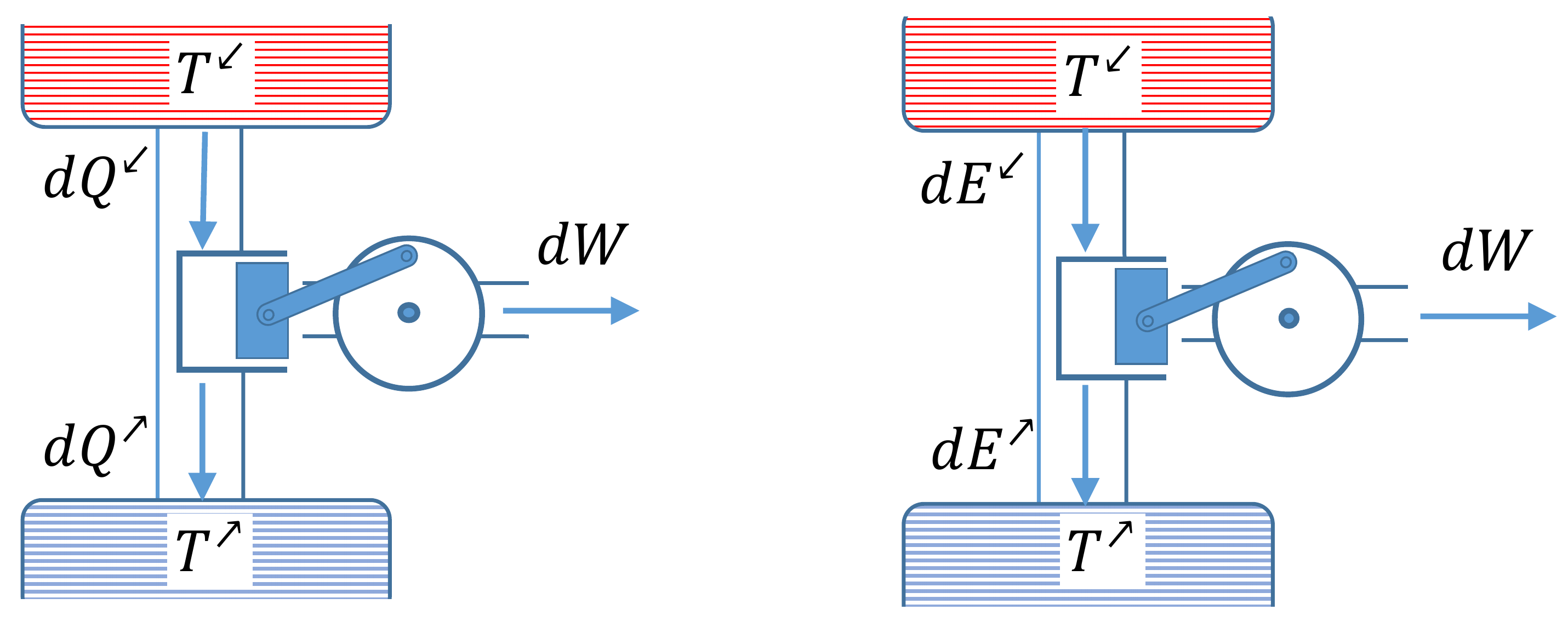}
	\end{center}
	\caption{The reformulation of the thermal engine as an energy engine.}
	\label{fig:eengine}
\end{figure}

\begin{figure}[t]
	\begin{center}
		\includegraphics[width=0.8\columnwidth]{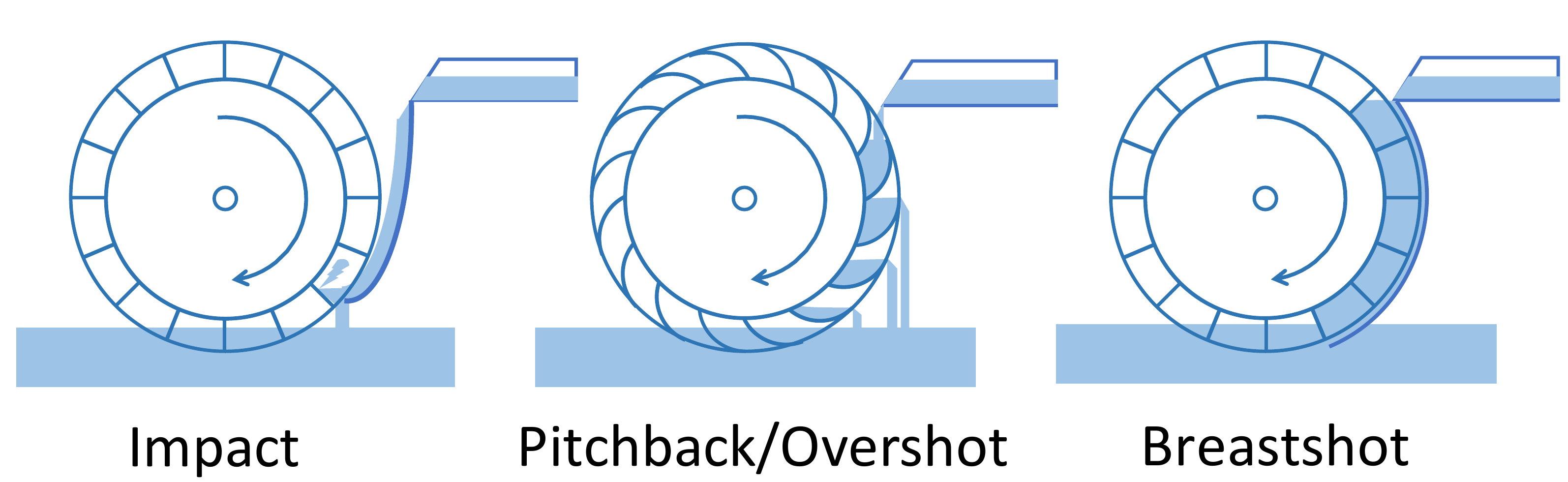}
	\end{center}
	\caption{Different ways of feeding water wheels. "Impact" symbolizes the transformation of 
		potential energy of water first into kinetic energy and then into work. "Pitchback" refers to leaking wheels, and "Breastshot" to a wheel with no leakage.}
	\label{fig:wheels}
\end{figure}

\begin{figure}[t]
	\begin{center}
		\includegraphics[width=0.6\columnwidth]{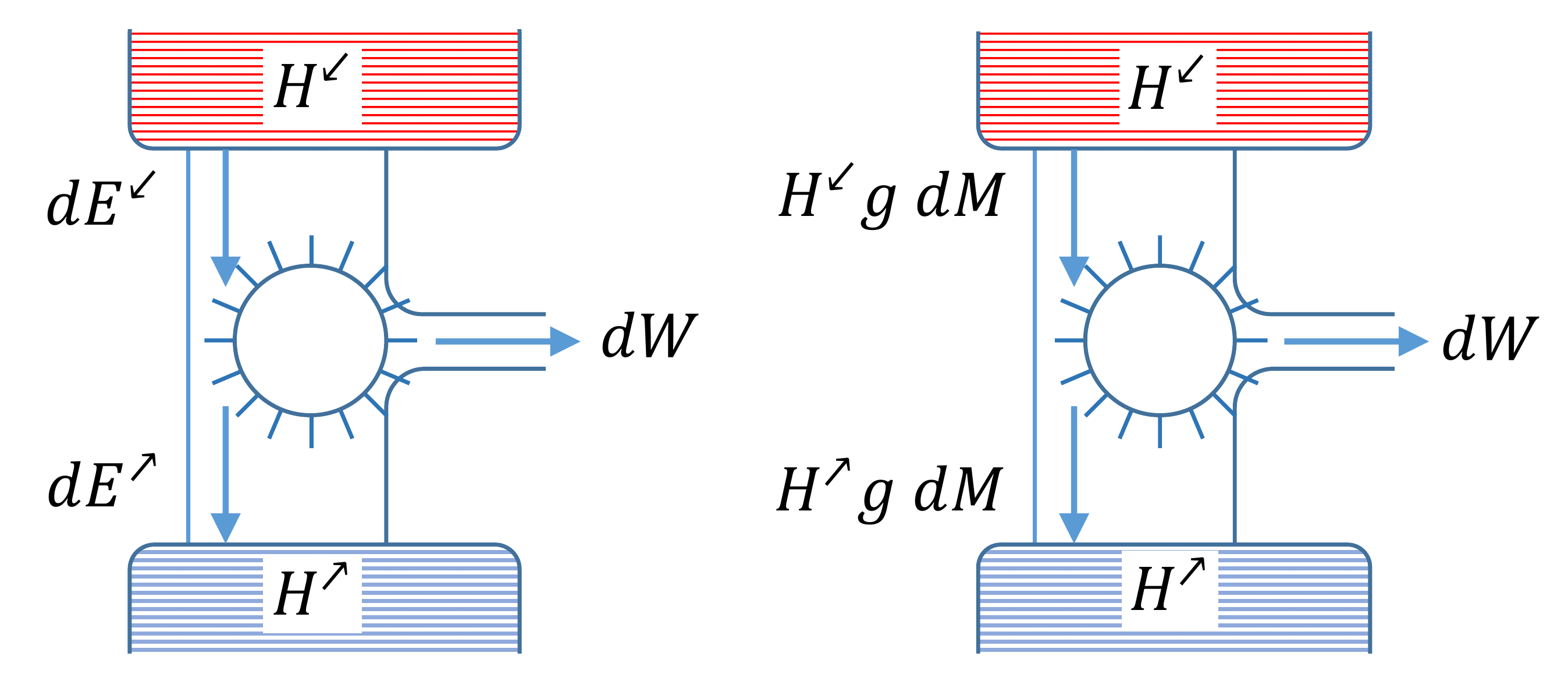}
	\end{center}
	\caption{A water wheel as an energy engine.}
	\label{fig:ewheel}
\end{figure}

For what concerns heat, I think that one should try to avoid the "$Q$" symbol, and simply speak of energy, as represented in Fig.~\ref{fig:eengine}. Notice that I also used the drawing of a piston instead of the circle to specify that there is no "continuity" between the heat reservoirs. 

But, in order to clarify that even if there were something similar to a fluid passing through the system, it would be different from the heat or the "caloric", let me introduce the water wheels, which have also the advantage of clarifying the concepts of reversibility and quasistaticity using a model which, in my opinion, is far more easy to understand with respect to thermal engines. 

\section{Water wheels} \label{sec:waterwheels}
Water wheels are known since ancient times, and studied in the engineering sense since the 1500, with the notable contribution of John Smeaton~\cite{smeaton} who, in 1759, experimentally determined the efficiency of different designs. Let us review the salient facts.  

A water wheel is a machine that continuously converts into an amount of work $d W$ the kinetic and gravitational energy of a certain mass $d m$  of water which enters with velocity $v^\swarrow$ at a height $H^\swarrow$ and which is discharged with velocity $v^\nearrow$ at a height $H^\nearrow$.  

The wheels can be built in many ways: they can only use the gravitational energy, with water speed $v^\swarrow=0$, like the \textit{overshot}  or \textit{breastshot} wheels in Fig.~\ref{fig:wheels}, or they can only exploit the kinetic energy as for example some wheels actuated from below, see \textit{impact},  again in Fig.~\ref{fig:wheels},

The question that puzzled the engineers of the past centuries, and in particular Smeaton, was: what is the shape and mode of action of a waterwheel that maximizes the instantaneous work $d W$? 

Let us consider again Fig.~\ref{fig:wheels}. In the \textit{impact} arrangement, the potential energy of the water is first transformed into kinetic energy and then this energy is used to move the wheel. The first conversion is extremely efficient: 100\% of potential energy can be converted into kinetic energy by a free fall. However, the second one corresponds to quite a waste of energy. There will be for sure turbulent motions, which will "dissipate" energy. 

Actually, the energy is still all there, according with the first principle of thermodynamics, the problem is that it was converted from an ordered motion to a microscopically disordered one. This fact may serve to illustrate the difference between the macroscopic kinetic energy and the temperature. 

Thus, it is much better to let water enter at lowest incoming velocity $v^\swarrow=0$ and obviously exit with the lowest possible velocity $v^\nearrow=0$. This is not always possible, like for instance for wheel operated by the stream of a river, called "undershot". 

In general,  the need of avoiding dissipation, and thus turbulence,  leads to a \textit{quasistatic} operation of the engines. However, this is not enough. 

The \textit{pitchback} wheel (Fig.~\ref{fig:wheels}) can be run quasistatically, but the leakage corresponds again to a waste of energy. It is  straightforward to show that the origin of dissipation is the same as in the first and second examples: in both cases there have been a transformation of potential energy into kinetic one, and then the conversion of at least a part of "coherent" kinetic energy into an incoherent one. 

Finally, the \textit{breastshot} wheel, if run quasistatically, corresponds to the maximum of the efficiency, which is mathematically defined below. It is also clear that this kind of wheel can be converted into a pump, feeding it with external work. 

It is easy to prove by contradiction that all reversible wheels have the same efficiency, otherwise one could pair a wheel with higher efficiency to one working as a pump, with the only result of being able to extract work without any loss of energy. 
Thus we see that also for water wheels the  maximum efficiency is  equivalent to reversibility. 

The \textit{breastshot} wheel is therefore like a Carnot engine working with two water reservoirs at two heights, as represented in Fig.~\ref{fig:ewheel}. 

Let now work out in details the analogy. It is clear that the equivalent of $d Q$, which is an energy exchange $dE$, is given by the potential energy of a quantity $dM$ of water entering or exiting at height $H$. The corresponding potential energy is 
\[
dE = Hg\,dM,
\]
where $g$ is the acceleration of gravity, considered constant since we assume  that the difference in height is small with respect the Earth radius.

The efficiency $\eta$ can be defined as 
\[
\eta = \frac{dW}{dE^\swarrow}.
\]

\begin{figure}[t]
	\begin{center}
		\includegraphics[width=0.4\columnwidth]{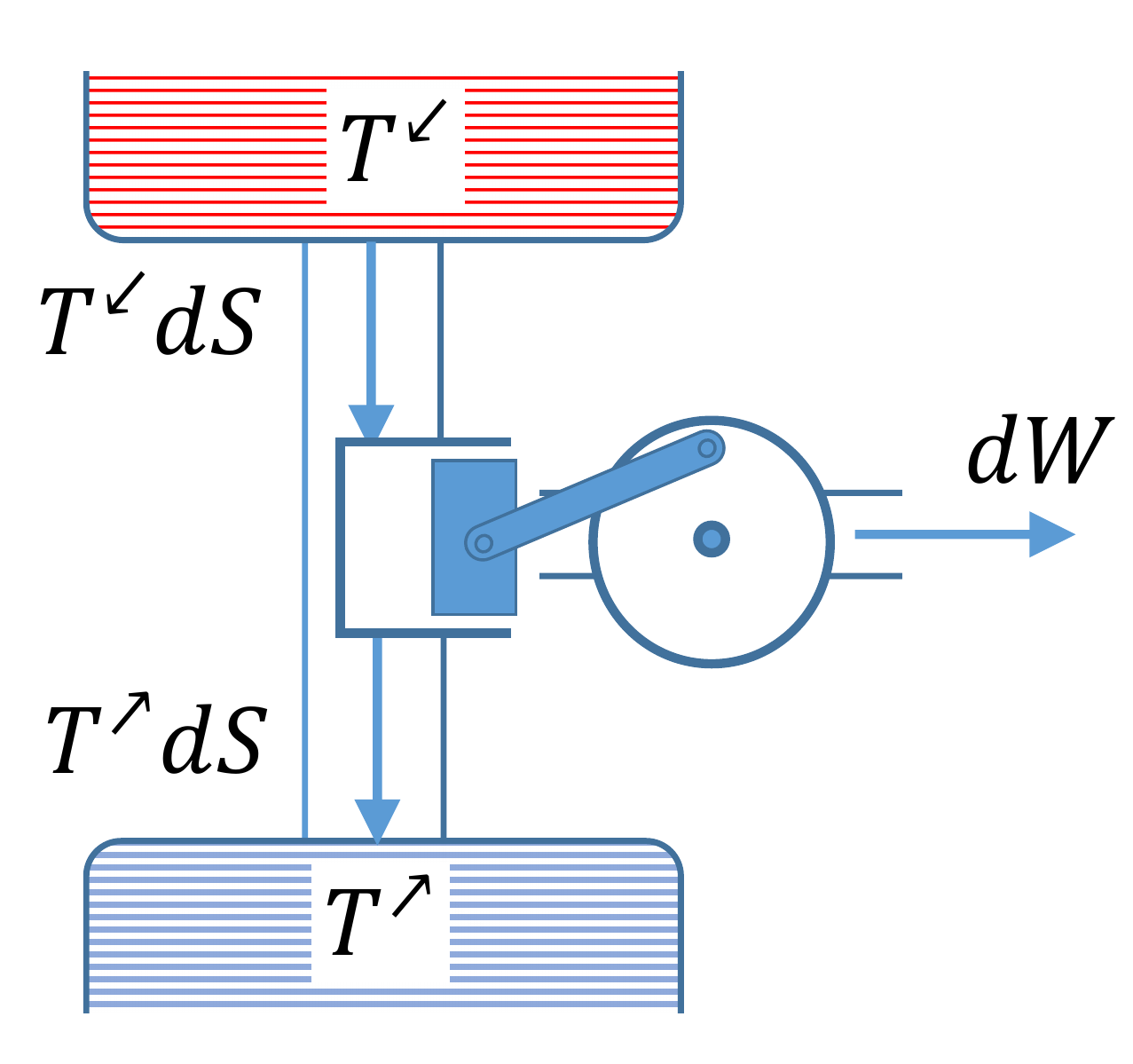}
	\end{center}
	\caption{The entropy engine.}
	\label{fig:sengine}
\end{figure}

Now, for a reversible water wheel, we have 
\[
dW= dE^\swarrow -dE^\nearrow = (H^\swarrow -H^\nearrow) g \,dM 
\]
and therefore
\begin{equation}\label{etah}
\eta = 1-\frac{H^\nearrow}{H^\swarrow}.
\end{equation}

It is evident that the height $H$ is the equivalent of the temperature $T$, and since we already said that the energy $dE=Hg\,dM$ is the equivalent of the heat $dQ$, we finally obtain what is the thermodynamic equivalent of the mass:
\[
g\,dM = \frac{dE}{H} \leftrightarrow \frac{dQ}{T} = dS.
\]
The mass of the fluid is nothing but the equivalent of the  thermodynamic entropy, which is conserved in reversible cycles, as shown in Fig.~\ref{fig:sengine}, and produced in non-ideal machines, see Sec.~\ref{sec:nonideal}.

It should be now clear why one cannot convert all the heat contained in a body into work. It would be like converting all potential energy of water into work, but in order to make the wheel turn one has to \textit{discharge} the water, like in thermal machine one has to discharge entropy. 

Before passing to the microscopic world, I have to stress that, while in mechanics the reference height for potential energy is arbitrary, one cannot use here the same approach since in this case also the efficiency $\eta$ of Eq.~\ref{etah} would be arbitrary. The same applies to the temperature for the efficiency of thermal engines, and is solved by the use of the absolute scale of temperature.
Similarly, we need to measure the heights starting from an absolute reference altitude, that can be assumed to be the centre of Earth.

Indeed, it is the lowest height at which a mass of water can be discharged, in principle. The problem is that all water "levels" up to sea one are  already occupied, and it makes little sense making holes just to fill them with water. Similarly, although the thermodynamic efficiency grows  the lower the temperature of the cold reservoir, we cannot in practice discharge heat at a temperature which is lower than that of the surroundings. 

\section{Going microscopic}\label{sec:micro}

\begin{figure}[t]
	\begin{center}
		\includegraphics[width=0.8\columnwidth]{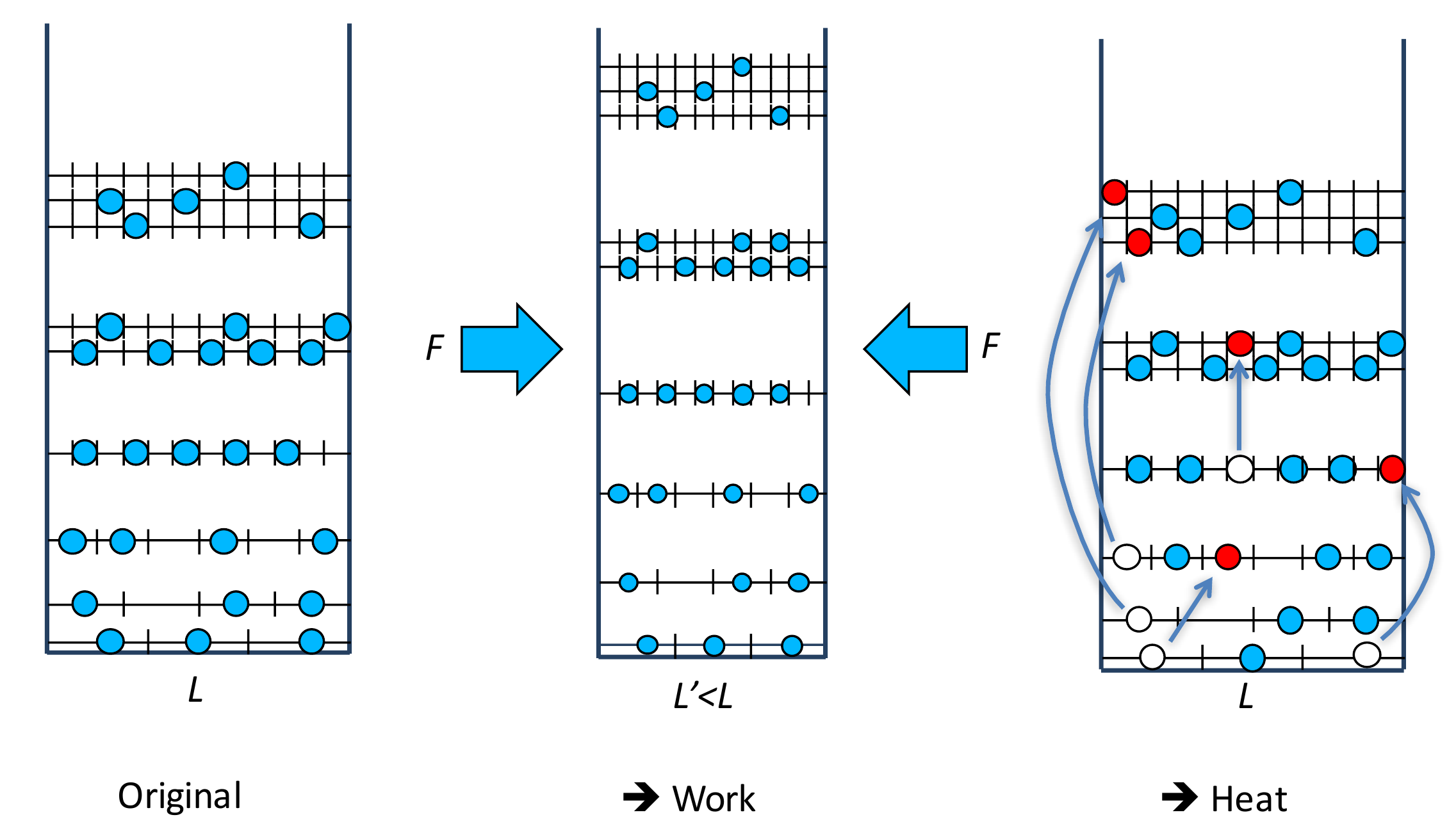}
	\end{center}
	\caption{Work and heat at the microscopic level.}
	\label{fig:workheat}
\end{figure}

In order to  offer an alternative view of the equivalence between water wheels and thermal engines, let me introduce some microscopic models. 

\subsection{Ideal gas}\label{sec:discrete}
For the Carnot engine, we need a model of an ideal gas. It is convenient to use a quantum approach, because it will make clear the dependence of the energy on the volume. Moreover, with a discrete model, the phase space is also discrete and numerable, so that the Shannon and Boltzmann entropies are mathematically well defined. 

We can use a simple model of a  perfect gas made by almost non-interacting and identical particles in a one-dimensional box of width $L$. As shown in the Appendix, the energy levels of the system are 
\[
E_n \propto \frac{h^2}{mL^2}n^2 ,
\]
where  what is important is that they increase their energy when the volume ($L$ in this case) is reduced. 

Let us indicate by $P_n$ the probability of finding a particle with energy $E_n$, and by $N$ the total number of particles. The number of particles with a certain energy is $N_n = NP_n$. The average energy per particle is
\begin{equation}
\label{eq:E}
\langle E\rangle  =\sum_n E_n P_n.
\end{equation}

\subsection{Work and heat at the microscopic level}
We are now ready to illustrate the equivalents of heat and work at the microscopic level~\cite{ngo}: a reversible work $dW$ is done on the system is a compression by an external force $F$, that slowly reduces the volume $L$. This corresponds to an increase of the energy without any change in the population, simply due to an increase  $dE_n$  of the energy levels, as illustrated in Fig.~\ref{fig:workheat}-Work. The opposite happens in an expansion. 

On the other hand, a transfer of energy at constant volume con only be obtained by a change in the population $d P_n$, as illustrated in Fig.~\ref{fig:workheat}-Heat. 

By differentiating Eq.~\eqref{eq:E} we obtain
\begin{equation}
\label{eq:dE}
dE =  \underbrace{\sum_n P_n dE_n}_{\text{\normalsize work}} +\underbrace{\sum_n E_n d P_n}_{\text{\normalsize heat}},
\end{equation}
where one can identify the two terms.

\subsection{Information entropy}

We can now define the entropy $S$, which is a functional of the probability distribution $P$. For two almost non-teracting systems $A$ and $B$, the entropy $S$  corresponds to the sum of the two entropies  $S(A+B)=S(A)+S(B)$, which expresses the extensive character of this quantity. 
We shall suppose for simplicity that the energy levels of the two systems are different, so that the total energy $E_n$ is given by $E_n=E_{n_A} + E_{n_B}$, without degeneracy. 
The probability $P$ for two independent processes is given by  $P_n(A+B)=P_{n_A}(A)P_{n_B}(B)$. It is therefore clear that the entropy has to do with the logarithm of the probability. Defining $S$ as the expectation value of it, we have
\[
S=-\sum_n P_n \log(P_n),
\]
and it can be easily checked that this form is extensive. 

The entropy defined in this way is also a measure of the information we have on the system, than we assume can take $\Omega$ different states. In the case of complete information, the system is in a given state $m$, and therefore $P_n=\delta_{nm}$ and  $S=0$. If we do not have any information,  $P_n = 1/\Omega$ and $S=\log(\Omega)$, which corresponds to the maximum value for the entropy.
 
By using a spreadsheet (with a formula to keep the normalization of the probability constant), one can  show that a flat distribution is indeed the one that maximizes the entropy. This corresponds to the Boltzmann case, in which one has, in the phase space, a certain number $\Omega$ of distinguishable configurations  at the same energy, i.e., a closed system. 

Notice that the "information" term is ambiguous. It can correspond to the information that could be obtained by carefully studying the system, which is an objective point of view, or it can be interpreted as the information that we think is sufficient to characterize the system, i.e., a subjective point of view~\cite{jaynes1,jaynes2}. In this latter case, the idea is the following: suppose that the information we put is sufficient to characterize the system, get the equilibrium distribution by maximizing the entropy and derive the value of observable quantities. If they agree with the experimental observation, the model is validated, otherwise more elements have to be added. 

In the case of a system in thermal contact with a reservoir, the hypothesis is that the average energy, which is constant due to the interaction with the reservoir, is the quantity that determines the probability distribution.  

\subsection{Equilibrium distribution}
By imposing  this constraints one can show that the distribution corresponding to the maximum of entropy  is indeed the Boltzmann distribution 
\begin{equation}
\label{eq:boltz}
P_n = \frac{1}{Z} \exp\left(-\frac{E_n}{T}\right),
\end{equation}
as can be checked again using a spreadsheet.

Clearly, one can mathematically get this expression by maximizing the entropy with the given constraints, by means of the Lagrange multipliers
\[
\frac{\partial}{\partial P_n}\left( -\sum_{n'} p_{n'} \log(P_{n'})-\lambda_1 \sum_{n'} P_{n'} -\lambda_2 \sum_{n'}E_{n'} P_{n'}\right) =0.
\]

\begin{figure}[t]
	\begin{center}
		\includegraphics[width=0.8\columnwidth]{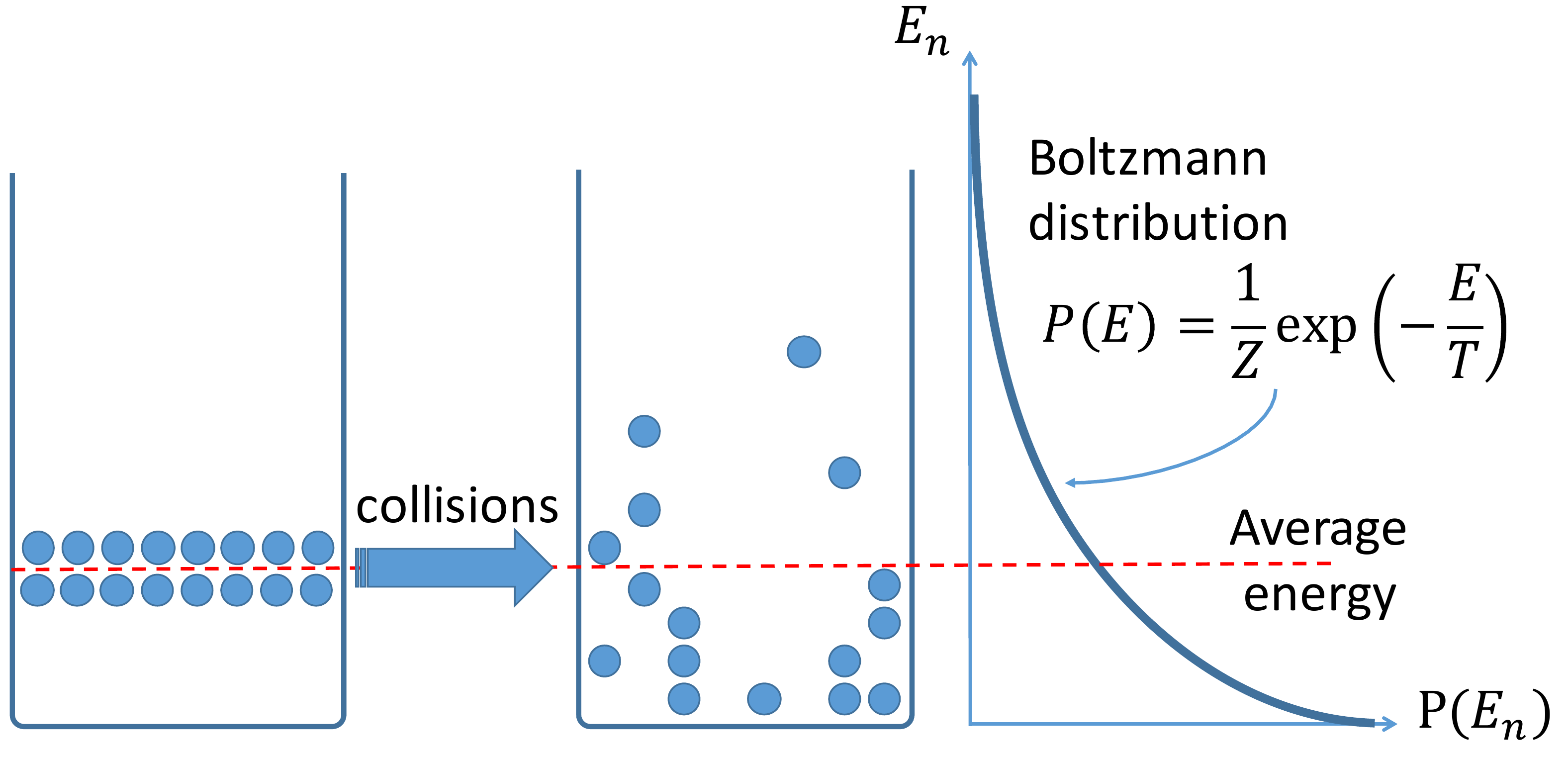}
	\end{center}
	\caption{Equilibrium distribution.}
	\label{fig:equilibrium}
\end{figure}

This is indeed a difficult point to illustrate to a classroom, but it is important to stress that the basic assumption of thermodynamics is that any system in contact with a thermal reservoir, will eventually reach the equilibrium distribution (see Fig.~\ref{fig:equilibrium}), regardless of the initial distribution, and that this distribution depends on just one parameter: $T$. 

\subsection{Temperature}
We can now identify this parameter. The variation  of the entropy $S$ is
\[
dS = -\sum_n \log(P_n) dP_n -\sum_n dP_n 
\]
where the last term is null due to the normalization condition $\sum_n P_n =1$. Inserting the Boltzmann distribution, Eq.~\eqref{eq:boltz}, one gets
\[
dS = \sum_n \left(\log(Z)+\frac{E_n}{T}\right)dP_n.
\] 
The first term is again zero due to the normalization condition of $P_n$. The second is just $dE/T$, Eq.~\eqref{eq:dE}, in the case of an isochoric transformation (no work). Thus
\[
T = \frac{dE}{d S} \qquad \text{or} \qquad dS=T dQ,
\]
since in this case the transfer of energy corresponds to the passage of heat.

As already said, the equilibrium distribution only depends on the  parameter $T$. This means that, regardless of the level distribution (i.e., of the material) the population at a given energy is always the same, for the same temperature.   

\begin{figure}[t]
	\begin{center}
		\includegraphics[width=0.8\columnwidth]{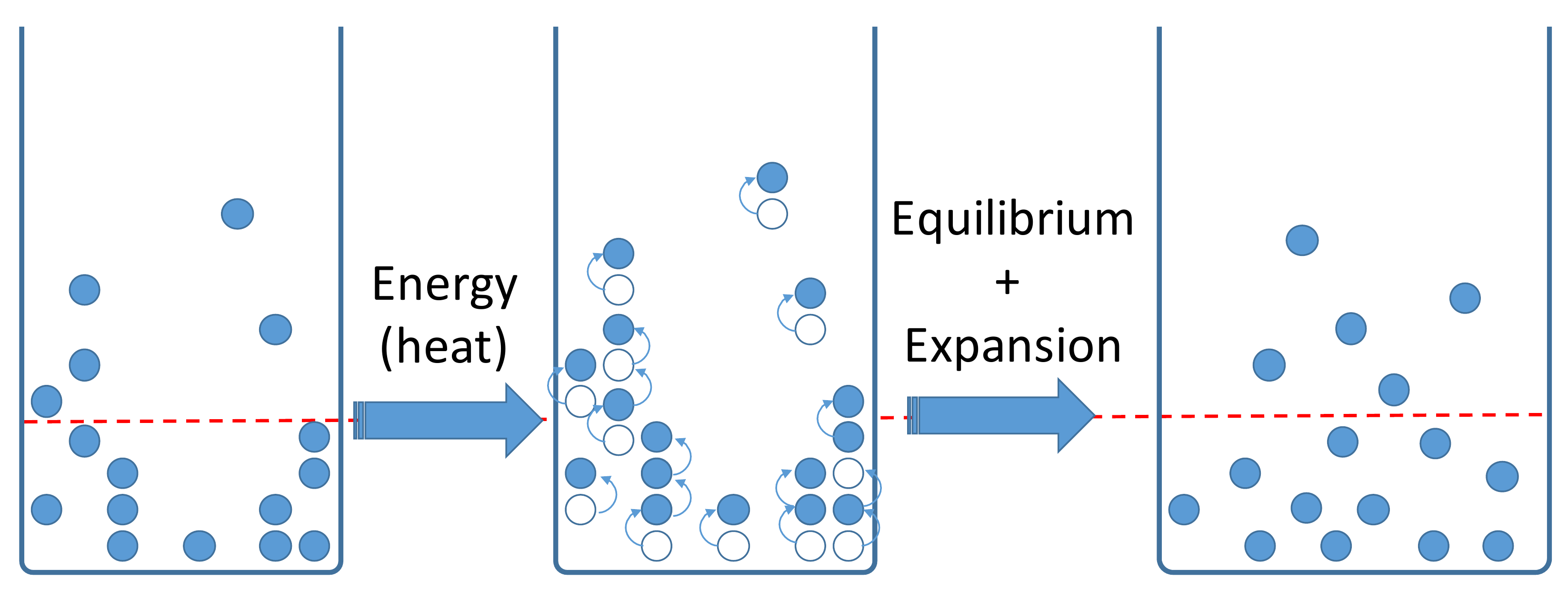}
	\end{center}
	\caption{Isothermal transformation, composed by a transfer ef energy (heat) followed by an expansion so that the average energy (temperature) remains constant. We have represented here the transfer of heat as jump of all particles, but it could be implemented by any other transformation of the population. }
	\label{fig:isothermal}
\end{figure}

The idea of equilibrium is the following:   take an arbitrary initial distribution, say, all particles in a given level or in two levels or in any other disposition. Particles can exchange energy among them, so the total energy  is kept constant, as is  the number of particles. The principle of maximum entropy says that the final probability distribution is the Boltzmann one, with an average energy per particle
\[
\begin{split}
\langle E \rangle &= \sum_n P_n E_n = \frac{1}{Z}\sum_n E_n \exp\left(-\frac{E_n}{T}\right)  \\
& =- \frac{1}{Z} \frac{dZ}{dT^{-1}} = T^2 \frac{d\ln(Z)}{dT}.
\end{split}
\]
In the case of the 1D box,  
\[
Z = \sum_n \exp\left(-\frac{E_n}{T}\right) = \sum_n \exp(-\frac{An^2}{T}),
\]
where $A=h^2/(2mL^2)$. 
Approximating the sum with an integral,
\[
\begin{split}
Z&=\sum_n \exp(-\frac{An^2}{T}) 
\simeq \int_0^\infty dx \exp\left(-\frac{Ax^2}{T}\right)  \\
&= \sqrt{\frac{T}{A}}\int_0^\infty dy \exp\left(-y^2\right)  = C\sqrt{T},
\end{split}
\]
where $C=\sqrt{\pi/(4A)}$. Since $\ln(Z) = (1/2) \ln(T)+ \ln(C)$, 
we have 
\[
\langle E \rangle = T^2 \frac{d\ln(Z)}{dT} = \frac{1}{2} T,
\]
that clearly represents the per particle contribution of the kinetic energy along one dimension. The important point is that of showing that the average energy is related to the temperature, by a constant (specific heat) that depends on the material and possible the temperature. 

\subsection{Carnot cycle}
Let us now try to describe the Carnot cycle at the microscopic level. We already illustrated the adiabatic compression/expansion. They simply consists in a change of the energy of levels, without any change in the population, at least in an ideal world. 

For the isothermal transformation, say an isothermal expansion, we have a simultaneous
transfer of energy and lowering of the energy of levels so to end with the same temperature as the beginning. This is illustrated in Fig.~\ref{fig:isothermal}. After the transfer of the energy the population relaxes to an equilibrium distribution with a different average energy, corresponding to a different temperature, and a different value of the entropy. The expansion phase brings again the temperature to the previous level, which is that of the heat reservoir. The reverse happens during a compression. 

So a heat engine works by lowering the energy of the particles composing the internal gas by means of an expansion, followed by an input of external energy, and thus of entropy, when in contact with the high-$T^\swarrow$ reservoir. After that, the energy of the gas is lowered  without any change in entropy by an expansion. The energy and the entropy are then expelled by means of a temporary rising of the energy of internal particles due to a compression, followed by the thermal contact with the low-$T^\nearrow$ reservoir. Finally the cycle closes by increasing the energy of the gas by a compression. 

\begin{figure}[t]
	\begin{center}
		\includegraphics[width=0.8\columnwidth]{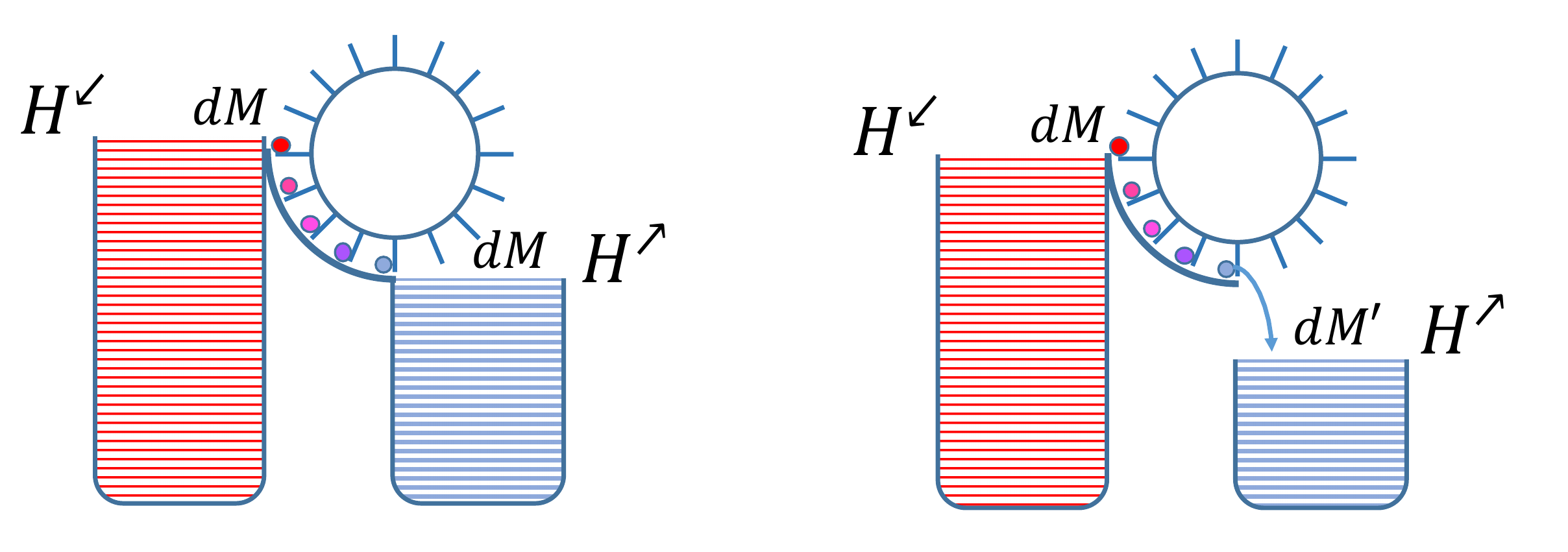}
	\end{center}
	\caption{Ideal and nonideal water wheels.}
	\label{fig:i-niwheel}
\end{figure}

But there is no need to suppose that the internal gas is always the same. If the reservoirs were also made of ideal gases, one could have part of this gas enter the system in the hight-$T^\swarrow$ contact during the expansion: the pressure of the gas of the reservoir is higher than that of the piston due to its higher temperature. Similarly, the gas can be expelled  into the low-$T^\nearrow$ reservoir, during the compression. In this case we have a mass of gas passing from the two reservoirs, a mass that transports energy and  entropy. 
This is indeed very similar to how a hydraulic wheel works,  see Fig.\ref{fig:i-niwheel}-left. 

Having identified the heat with the energy transferred and the height with the temperature, it is apparent that the entropy, $d S=d Q/T$ is the analogue of the mass of water $d M$ passing through the reversible water wheel, apart from the constant $g$. 

Thus, as a hydraulic wheel works by means of water which \textit{falls} from a height $H^\swarrow$ to a height $H^\nearrow$, so the thermal machines operate by means of an entropy stream that \textit{falls} from a temperature  $T^\swarrow$ to a temperature $T^\nearrow$. It is also evident the need of dissipating the heat at a temperature $T^\nearrow$: as water cannot disappear, so the entropy cannot be destroyed.

\begin{figure}[t]
	\begin{center}
		\includegraphics[width=0.8\columnwidth]{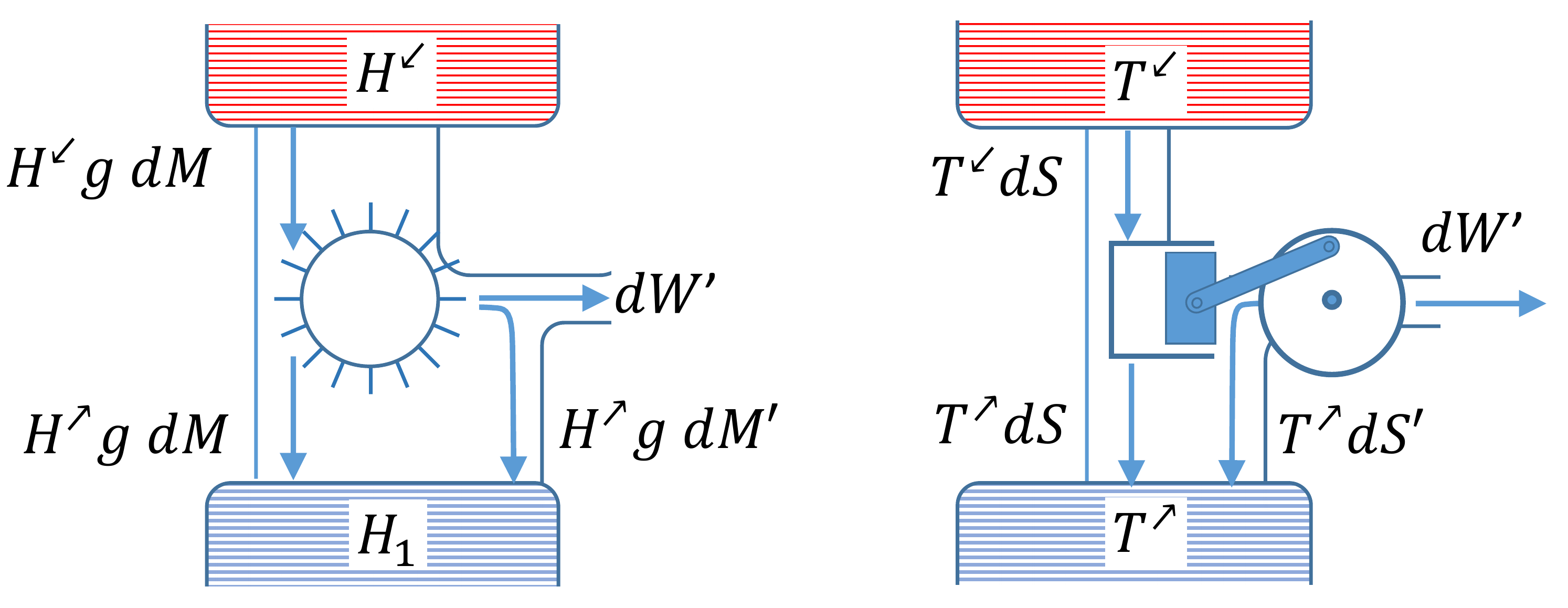}
	\end{center}
	\caption{Comparison between non-ideal water wheel and a non-ideal thermal engine}
	\label{fig:nonideal}
\end{figure}

\section{Non-ideal engines and entropy increasing}\label{sec:nonideal}
Real heat engines, however, are not reversible. Analysing the causes of irreversibility, that may be due to finite thermal jumps, turbulence or friction, it is evident that they have the same origin as those that \textit{transform} a reversible waterwheel into a non-ideal one, see Fig.~\ref{fig:i-niwheel}. 

Unlike the mechanical case, however, a non-reversible thermal machine outputs more entropy than what entered, while the waterwheels always discharge the same amount of water. The difference lies in the fact that the water discharged by a non-reversible wheel contains more energy than that discharged by a reversible one: either the water exits with a finite speed, or exhibits a turbulent motion, or it is warmer due to internal friction. At the end, all this wasted energy goes into heat. But while the heat added to the wasted water does not increase the discharged  mass, the heat added to the exhaust heat in thermal engines increases the discharged entropy. 

Fortunately, we can make use of the Einstein relation between mass and energy~\cite{mc2},
\[
dE=c^2 dM \qquad \Longrightarrow \qquad dM = c^{-2} dE.
\]
The energy wasted for the leakages or the dispersion of kinetic energy is still contained into water, and corresponds to an increase of discharged mass. Indeed, hot water has more mass than cold water, see Ref.~\cite{carlip}. So, a non-ideal water wheel is discharging more mass than the minimum one, as shown in Fig.~\ref{fig:nonideal}-left. 

 The mass increment is $\delta M = \Delta M\, C_V\Delta T/c^2$, where $C_V=4186\; \si{\joule \per \kilo\gram\per\kelvin}$  is the  specific heat of water, $\Delta T$ is the difference between the temperature of the water discharged and that of the environment,  and $c=3\cdot 10^8\;\si{\metre\per\second}$  is the speed of light. Since the latter  is so large, this mass increment is  a mere $4\cdot10^{-14}~\si{\kilo\gram}$ for each kilogram $\Delta M$ and degree difference $\Delta T$. 
 
\begin{figure}[t]
	\begin{center}
		\includegraphics[width=0.6\columnwidth]{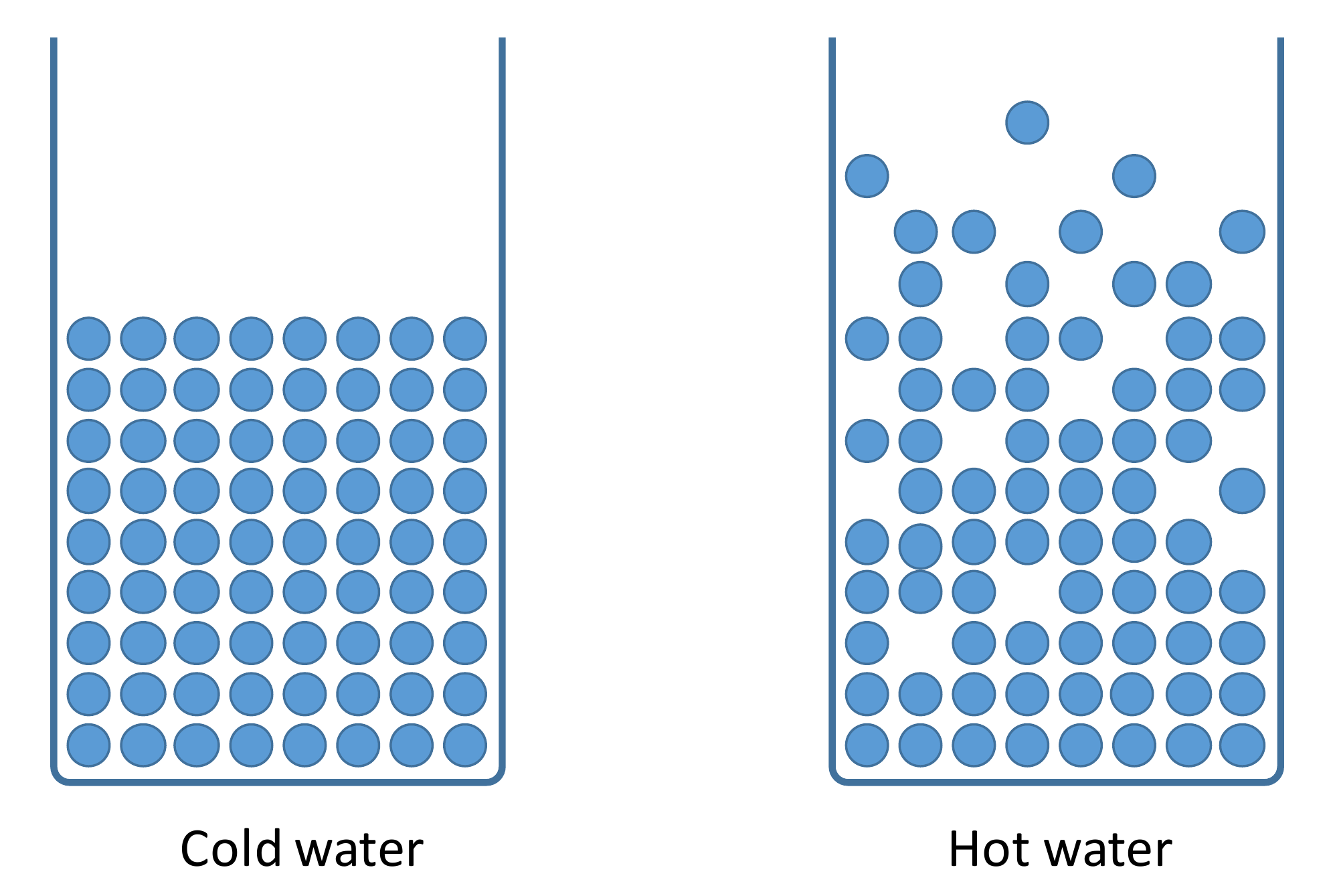}
	\end{center}
	\caption{Cold and hot water in the energy scale. The cold water is represented by balls that pile up due to the electric repulsion (and the Pauli exclusion principle). The hot water distribution is obtained by moving some ball higher in energy, and therefore its average energy is higher than the cold water one. }
	\label{fig:hot}
\end{figure}

\subsection{Hot water} \label{sec:mixup}

What does happen if we mix-up things, having hot water move a waterwheel? It is something like the example of the  Carnot engine moved by ideal gases that is entering and exiting the piston, as discussed in the previous Section.  Let us suppose that the higher reservoir is hotter that the lower one. 

The entering water has more energy (see Fig.~\ref{fig:hot}) and thus more mass, due to its higher temperature, with respect to the bare one. If it is discharged with unchanged temperature at the lower height by a reversible waterwheel, we still have a waste of energy, and thus an increase of entropy,  since its temperature  is higher than that of the surroundings. It is again the case represented in Fig.~\ref{fig:i-niwheel}-left.

The reversible waterwheels in this case is not the most efficient engine, since it cannot extract the energy difference contained in the thermal motion. Actually, it is no more reversible, since it cannot warm the water when acting as a pump.

\section{Third law of thermodynamics (for water wheels)}\label{sec:thirdlaw}
It is also possible to introduce the third law of thermodynamics using the hydraulic analogy. Since the entropy corresponds to the mass of water,  the temperature corresponds to the height with respect to the center of the Earth, and a heat reservoir corresponds to a water container.  The equivalent of an isolated body is a \textit{planet} composed of water only. 

Assuming that the water continues to be incompressible at all pressures (constant density), the volume $V$ corresponding to a certain height $H$  is $V=(4/3) \pi H^3$, and then the "entropy",  which corresponds to  the mass, i.e.,  density per volume, vanishes when $H\rightarrow 0$,   \textit{i.e.}, in the limit of zero temperature.

We can also define the analogue of the heat capacity $C$, which defines the increment of the temperature of a body when we put energy in: 
\[
C = \frac{\Delta Q}{\Delta T}.
\]

For the hydraulic analogy it is to better define the entropic capacity 
\[
\chi = \frac{\Delta S}{\Delta T}; \qquad C=T\chi.
\]

The hydraulic analogue of the transfer of the entropy to a heat reservoir consists in pouring water into a basin. Then, the hydraulic equivalent of $\chi$  is obtained by replacing $\Delta S$ by the mass change $\Delta M$  of the water contained in the basin, and $\Delta T$ by the variation of the height. Since water is incompressible, $\Delta M = \rho \Delta V$, where $\rho$  is the density and $V$  the volume. The increase $\Delta V$  of the latter is given by the product of the surface $A$ by the variation of the height $\Delta H$. 

So, apart from constants, $\chi$, which corresponds to the ratio $\Delta V/\Delta H$, is simply the surface  $A$ of the basin (for that height). Note that this derivation does not imply that the surface section is constant, actually it changes with the height. 

Thus, the analogy that is sometimes drawn in calorimetry between the heat capacity $C$ and the \textit{base surface} of a container, and between the temperature $T$ and the height level of the \textit{caloric} content of the body can be tightened by defining the amount of entropy, instead of the amount of heat, contained in the body, and specifying as the entropic capacity $\chi$  varies with the temperature. 

Notice that the heat capacity $C=\chi T$  does not corresponds to the volume of the basin, because this would imply a \textit{cylindrical} shape of it, impossible to be extended to the center of the Earth. Instead, $C$ is simply the product between the height of the free surface of water (entropy) and the surface of the basin itself. 

If we consider a planet as the hydraulic-gravitational equivalent of a thermally isolated body, it is easy to show that the incompressibility of water implies that the surface $\chi$ decreases with the height $H$, as the quantum ``thermal capacity'' decreases with the temperature. 

In particular, at low heights, $C\propto H^3$ , which corresponds to the specific heat of phonons in the Debye approximation~\cite{goodstein}. 

\section{Conclusions}\label{sec:conclusions}

The analogy between water wheels and heat engines allows to introduce the concept of entropy, which is always rather difficult to digest for the students, in a more visual way, and also to show that many of thermodynamics considerations do not only apply to thermal engines, but to machines in general. 

For instance, this analogy can lead to considerations about efficiency and the rational use of energy. The most striking example is that of an electric heater. If we measure the efficiency of an electric heater using the radiated thermal energy as the work produced, we have a efficiency almost equal to one. But if we consider that the electricity has been produced with a thermal power plant that discharges as \textit{waste} heat a lot of energy at almost the temperature of the heater, and that the conversion of energy and its transportation is quite inefficient, one obtains that the actual efficiency is quite low. Let us try to follow an analogous hydraulic process: we have to \textit{produce} water (heat) at a certain height (temperature), using electric power. The only (theoretical) way we have of doing so is to use nuclear processes and directly convert the energy into matter. 

The senselessness of the proceedings is immediately obvious, thinking that the production of the electricity needed by our nuclear laboratory requires maybe a hydroelectric power plant, that probably discharges used water at the desired height. It is much more efficient to use a pump to raise the water, as well as in the thermal case it is more efficient to use a heat pump to rise the temperature. And in both cases, when feasible, it is still more efficient to use the waste energy (or water), at the same temperature/height. 

In any case, I think that the proposed analogy can be fruitful exploited in physics course, making clear the distinction between energy/heat and entropy. Moreover, the microscopic models presented can be used to let pupils play and propose simple mechanical mechanisms for representing the working of  thermal, hydraulic or mixed engines.

\section*{Acknowledgements}
I would like to acknowledge fruitful discussions with Angelo Baracca.

\section*{Appendix}
It is relatively easy to derive the spectrum of a one-dimensional quantum well using the uncertainty principle. Let's assume for granted that for quantum particles the relation between the precision in position $\Delta x$ and in that in velocity $\Delta v$ is
\begin{equation}\label{hbar}
\Delta x \cdot m \Delta v \simeq h,
\end{equation}
where $h$ is of order of the Planck constant (I shall neglect constant factors in the following). 

If a particle is confined in a one-dimensional box of width $L$, clearly $\Delta x=L$. This means that  
\[
m \Delta v \simeq \frac{h}{L}.
\]
Since the uncertainty on the velocity does not depend on the velocity itself, we can assume that the velocity can only take discrete values, spaced by this quantity. Thus
\[
m  v_n \simeq \frac{nh}{L},
\]
with $n=1,2,\dots$. Now, the energy for this particle is just kinetic energy, and therefore 
\[
E=\frac{1}{2}mv^2, 
\]
and it is therefore quantized:
\[
E_n = \frac{1}{2} m v_n^2 = \frac{ h^2n^2}{2mL^2}.
\]
This expression is remarkably similar to the exact one
\begin{equation}\label{Ewell}
E_n = \frac{ h^2n^2}{8mL^2} . 
\end{equation}

\begin{figure}[t]
	\begin{center}
		\includegraphics[width=0.8\columnwidth]{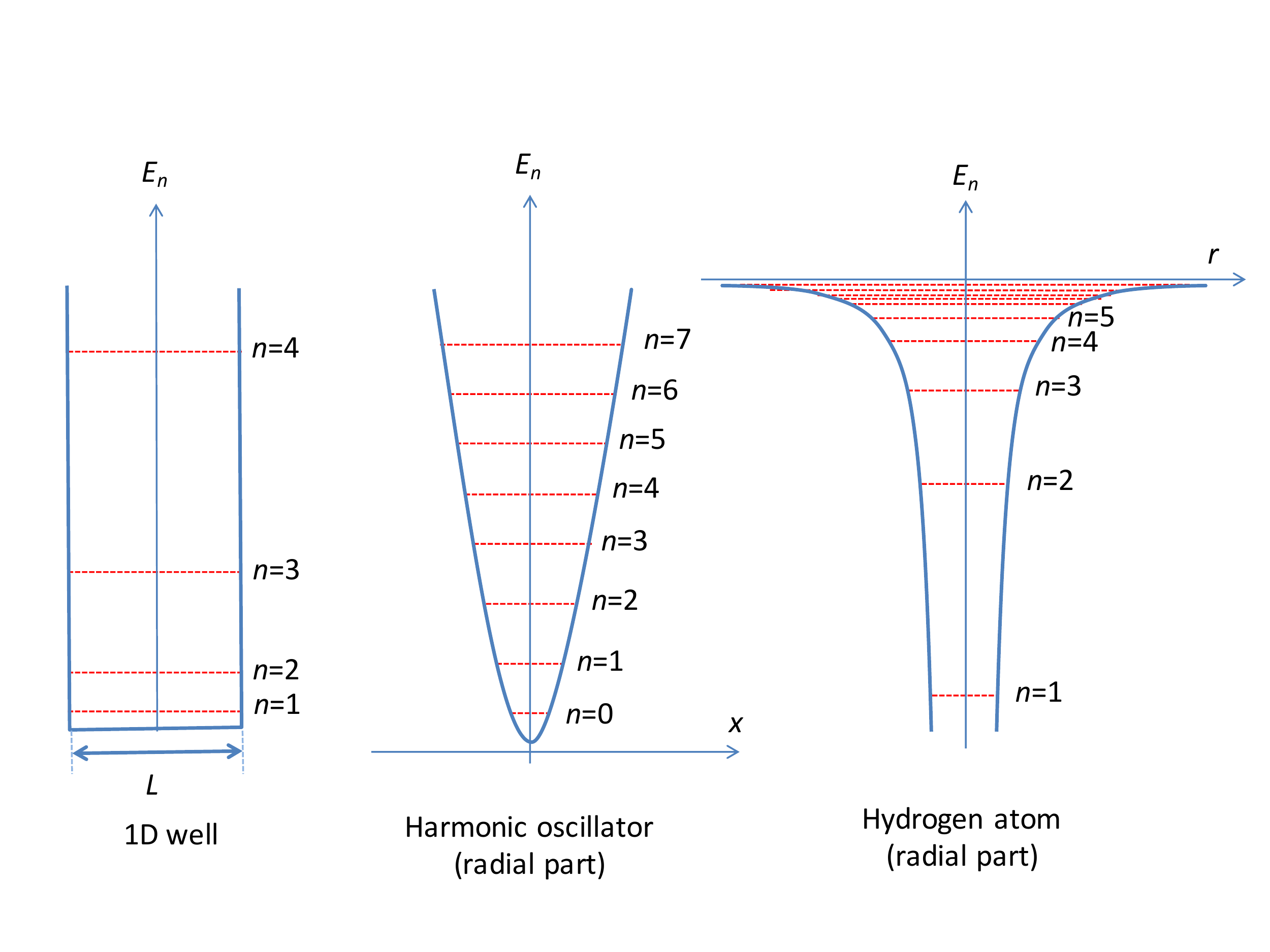}
	\end{center}
	\caption{Energy spectra of a 1D well, harmonic oscillator and hydrogen atom}
	\label{fig:spectra}
\end{figure}

What is important here is to remark that the spacing between levels depends inversely on the mass of the particle and on the width of the box. For macroscopic particles in a macroscopic box, the levels are essentially continuous, any velocity is allowed. 

Our ideal gas will be composed by a set of almost non-interacting particles in a well, where the "almost" is due to the fact that they have to thermalise, so a little interaction is necessary.

One can obtain also the shape of the spectrum for the harmonic oscillator. In this case the energy is given by 
\[
E = \frac{1}{2}mv^2 + \frac{1}{2}k x^2.
\]
The energy alternates between all-kinetic (for $x=0$) to all-potential ($v=0$). Using a semi-classic approach, i.e., the correspondence principle, that will be  accurate for high  energies, one can estimate that the effective width of the "cage" (equivalent to the width of the box in the previous case) is of order 
\[
L \simeq \sqrt{\frac{E}{k}}
\]
Inserting this expression in place of $L$ in Eq.~\eqref{Ewell}, we get
\begin{equation}
E_n \simeq  n\hbar \omega,
\end{equation}
with 
\[
\omega = \sqrt{\frac{k}{m}},
\]
and disregarding the numerical factors. Actually, if one uses the average square position for the classic harmonic oscillator one gets a quite accurate result. 

A set of almost non-interacting quantum oscillators represents  the model of a solid proposed by Einstein~\cite{einsteinsolid,goodstein}. The effects of an diminishing  of the volume can be made to correspond to an increase of the rigidity of the spring $k$, since the stronger the spring, the smaller the width of the  oscillations.

Just as an almost useless demonstration of virtuosity, using the same approach it is possible to obtain the form of the spectrum of the harmonic oscillator
\[
E = \frac{1}{2}m v^2 - \frac{e^2}{4\pi \varepsilon_0 r}
\]
using 
\[
L \simeq  -\frac{e^2}{4\pi \varepsilon_0 E}
\]
one obtains an expression that differs from the exact value
\[
E_n = - \frac{m e^4 }{8 h^2 \varepsilon_0^2 n^2}
\]
only for numeric constants. 

So far we have obtained the spectrum of a one-dimensional quantum well or harmonic oscillator. 
In two or three dimensions, the kinetic energy depends on the sum of the three components, so 
one can easy illustrate (using a sheet od squared paper) that the number of levels (degeneracy) of a given energy increases with the energy.


\begin{thebibliography}{99}
	
\bibitem{kesidou} S. Kesidou, R. Duit,  \textit{Students' conceptions of the second law of thermodynamics -- an interpretive study}, J. Res. Sci. Teach. \textbf{30}, 
	85 (1993). doi:10.1002/tea.3660300107
	
\bibitem{atkins} P.W. Atkins, \textit{The Second Law}, Scientific American Books (H. Freeman and Company, New York 1984). 	

\bibitem{kincanon} E. Kincanon, \textit{How I teach the second law of thermodynamics}, Physics Education \textbf{48}, 91 (2013). doi:10.1088/0031-9120/48/4/491

\bibitem{styer} D.F. Styer, \textit{Insight into entropy}, Am. J. Phys. \textbf{68},  1090 (2000). doi:10.1119/1.1287353

\bibitem{schoepf} D.C. Schoepf, \textit{A statistical development of entropy for the introductory physics course}, Am. J. Phys. \textbf{70}, 128 (2002). doi:10.1119/1.1419097

\bibitem{Leff1} H.S. Leff, \textit{Removing the Mystery of Entropy and Thermodynamics -- Part I}, The Physics Teacher \textbf{50}, 28    (2012). doi:10.1119/1.3670080
\bibitem{Leff2} H.S. Leff, \textit{Removing the Mystery of Entropy and Thermodynamics -- Part II}, The Physics Teacher \textbf{50}, 87  (2012). doi:10.1119/1.3677281
\bibitem{Leff3} H.S. Leff, \textit{Removing the Mystery of Entropy and Thermodynamics -- Part III}, The Physics Teacher \textbf{50}, 170    (2012). doi:10.1119/1.3685118 
\bibitem{Leff4} H.S. Leff, \textit{Removing the Mystery of Entropy and Thermodynamics -- Part IV}, The Physics Teacher \textbf{50}, 215    (2012). doi:10.1119/1.3694071 
\bibitem{Leff5} H.S. Leff, \textit{Removing the Mystery of Entropy and Thermodynamics -- Part V}, The Physics Teacher \textbf{50}, 274    (2012). doi:10.1119/1.3703541

\bibitem{kolmogorov} A.N. Kolmogorov,\textit{ New Metric Invariant of Transitive Dynamical Systems and Endomorphisms of Lebesgue Spaces}, Doklady of Russian Academy of Sciences,  \textbf{119}, 861 (1958).

\bibitem{sinai} Y.G. Sinai, \textit{On the Notion of Entropy of a Dynamical System}, Doklady of Russian Academy of Sciences \textbf{124} , 768 (1959).

\bibitem{ott} E. Ott, \textit{Chaos in dynamical systems} (Cambridge University Press 1994) pp. 138--144. ISBN:0-521-43799-7

\bibitem{carnot} S. Carnot, \textit{Reflections on the Motive Power of Heat} second edition, (John Wiley \& SONS, New York 1897). 

\bibitem{baracca} A. Baracca, \textit{Una proposta di introduzione storica intuitiva ai concetti entropici}, La Fisica Nella Scuola \textbf{ 41}, 160 (2008). 

\bibitem{smeaton} J. Smeaton, \textit{An Experimental Enquiry concerning the Natural Powers of Water and Wind to turn Mills and Other Machines, Depending on a Circular Motion}, Phil. Trans. \textbf{51} (1759). doi:10.1098/rstl.1759.0019

\bibitem{ngo} C. Ng\^o, H. Ng\^o, \textit{Physique statistique} (Dunod, Paris 2008) ISBN:978-2100501588


\bibitem{jaynes1} E.T.  Jaynes,  \textit{Information Theory and Statistical Mechanics I}  Phys. Rev. \textbf{106}, 620 (1957). doi:10.1103/PhysRev.106.620

\bibitem{jaynes2} E.T.  Jaynes,  \textit{Information Theory and Statistical Mechanics II} Phys. Rev.   \textbf{108} 171 (1957). doi:10.1103/PhysRev.108.171


\bibitem{mc2} F. Rohrlich, \textit{An elementary derivation of $E=mc^2$}, Am. J.  Phys. \textbf{58},   348 1990. doi:10.1119/1.16168

\bibitem{carlip} S. Carlip, \textit{Kinetic Energy and the Equivalence Principle} (also available as preprint: arXiv:gr-qc/9909014), Am. J. Phys. \textbf{66}, 409-413,(1998). 
doi:10.1119/1.18885  

\bibitem{goodstein} D.L. Goodstein, \textit{States of Matter} (Dover, New York 1985). ISBN:9780486495064


\bibitem{einsteinsolid}  See for instance the \textit{Einstein Model of a Solid} on hyperphysics, \url{http://hyperphysics.phy-astr.gsu.edu/hbase/therm/einsol.html}


\end{thebibliography}
\end{document}